\documentclass[slac_one]{revtex4}
\usepackage{graphicx}
\usepackage{fancyhdr}
\begin{document}

\title{Multiplicity and  momentum distributions
of hadrons in deep inelastic scattering at HERA energies} 

%

\author{T.~Tymieniecka (on behalf of the H1 and ZEUS Collaborations)}
\affiliation{University of Warsaw, Warsaw and University of Podlasie, Siedlce, Poland}

\def\epem{{e^+e^-}}

\begin{abstract}
The universality and scaling behaviour of hadronisation have been
investigated in semi-inclusive
neutral current deep inelastic $ep$
scattering; specifically
the multiplicity and scaled momentum spectra of charged hadrons
have been measured and analysed. 
The measurements are performed in the current region
of the Breit frame, as well as in the current
fragmentation region of the hadronic centre-of-mass frame.
The data  collected
at the HERA ep collider by the H1 and ZEUS experiments
are compared with  similar measurements obtained
in $e^+e^-$ annihilation and
with previous $ep$ measurements as well as
with leading-logarithm parton-shower  predictions.
\end{abstract}

\maketitle

\thispagestyle{fancy}


\section{INTRODUCTION}

The study of parton fragmentation and hadronisation
processes provides valuable insight into the non-perturbative
regime of Quantum Chromodynamics (QCD).
These processes are studied by the H1 and ZEUS Collaborations 
at  HERA energies (about 300~GeV in the $ep$ centre-of-mass frame)
using the semi-inclusive charged particles produced 
in neutral current deep inelastic scattering (DIS).
Two quantities describe the properties of hadron production: 
the invariant mass of the hadronic system $W$ and
the virtuality $Q^2$ of the exchanged boson. 
The best way  to study hadronisation is to
look from the point of view of the exchanged boson colliding
with the incoming proton. In their centre-of-mass frame, 
called also the hadronic centre-of-mass frame (HCM),
half of energy, $W/2$, is carried out  by particles which have absorbed
the boson and the other half  by particles from the proton remnant. 
Na\"{i}vely one would expect  particle production
in the proton remnant region to be similar to particle production  
in $pp$ collision. Among the particles which have absorbed the exchanged
boson, a part of them is expected to have similar features
to the ones observed in one hemisphere of $e^+e^-$ annihilation.
The separation  is easily done in the Breit frame.
In this frame the exchanged boson of $Q/2$ energy is absorbed 
by the scattered quark which converts into particles in the region
called the current region of the Breit frame.  
Thus, depending on the rapidity of the emitted hadron its feature
can be similar either to the one produced in $e^+e^-$ annihilation
or in $pp$ collision.
There is a region of rapidity in between the origin of the Breit frame
and the HCM frame, called the central region,
 which is unique in its feature
for $ep$ scattering. This region is dominated by gluon emission.
In asymmetric collisions such as  $ep$~scattering leakage between  regions
modifies this simplified picture.

The current region of the Breit frame is populated mainly
by hadrons coming from  the zero-order QCD
process $\gamma q \rightarrow q$ 
and the first-order QCD process $\gamma q \rightarrow g q$.  
The other first-order QCD process is the boson-gluon
fusion (BGF) $\gamma g \rightarrow q \bar{q}$
with hadrons populating the edge of the current and
central region. 
This process  with its higher orders 
 as well as the initial state gluon radiation
do not contribute to hadron production in $e^+e^-$
and are the main source of differences  between hadron 
features in $e^+e^-$ and $ep$.
The depopulation of the current region depends~on~$Q^2$.

\begin{figure*}[t]
\centering
\includegraphics[width=90mm]{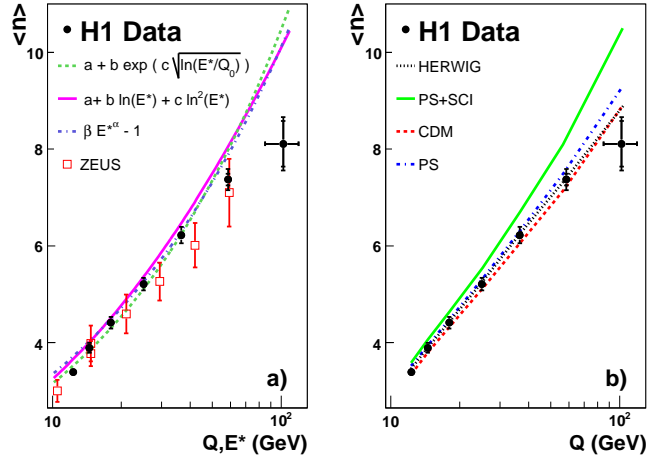}
\caption{The average charged multiplicity $\langle n \rangle$
as a function of $Q$ for DIS is compared with  a parameterisation of  
the $e^+e^-$ data~\cite{ee} (left) and with
MC predictions (right).
} \label{obraz1}
\end{figure*}

In this short review  fragmentation features in $ep$ scattering
 are presented
in form of the scaled momentum distributions measured 
in the current region of the Breit frame \cite{Aaron:2007ds,proc:zeusff:2007}
and of the average charged multiplicities \cite{Aaron:2007ds,zeus}.
The sensitivity to the energy scale definition is discussed for scales
given by $Q$, $W$ and a sum of the emitted energies.
The data are compared with predictions from Monte Carlo (MC) event generators.
Two  models,  ARIADNE 4.12~\cite{cpc:71:15} 
and LEPTO-MEPS 6.5.1~\cite{cpc:101:108}, are considered to describe
the DIS processes. In ARIADNE  the parton cascade is modelled
with the colour-dipole model whereas  LEPTO  incorporates 
the leading-order matrix element with some next-to-leading (NLO) correction
implemented in the parton shower. A soft colour interaction model (SCI) implemented
in LEPTO is also considered. For these event generators,
 hadronisation is performed
by the Lund string model~\cite{prep:97:31}.  In addition
HERWIG is discussed due to its cluster hadronisation model~\cite{herwig}
although this model did not describe similar data in the past.

\begin{figure*}[t]
\centering
\includegraphics[width=62mm]{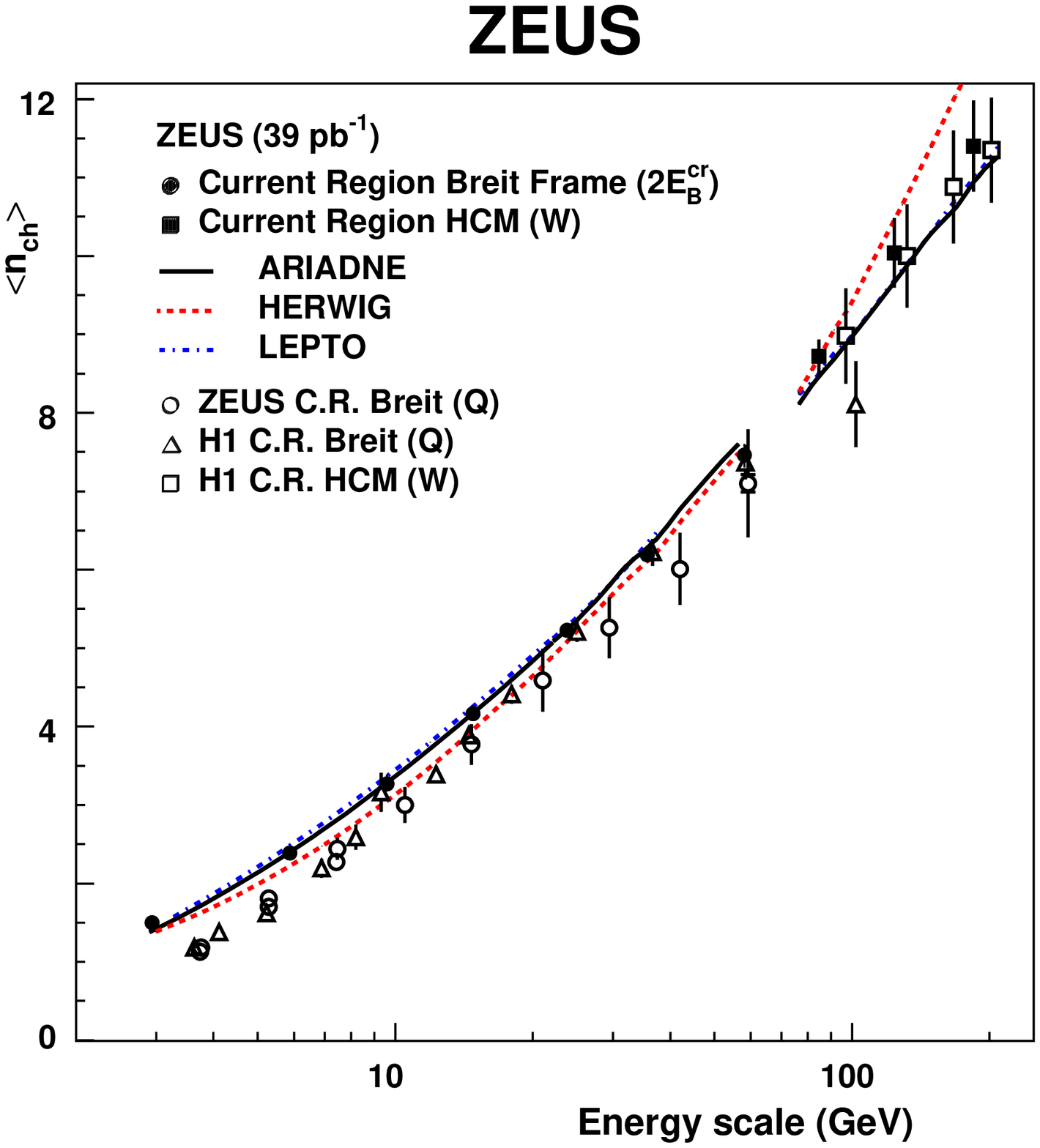}
\includegraphics[width=62mm]{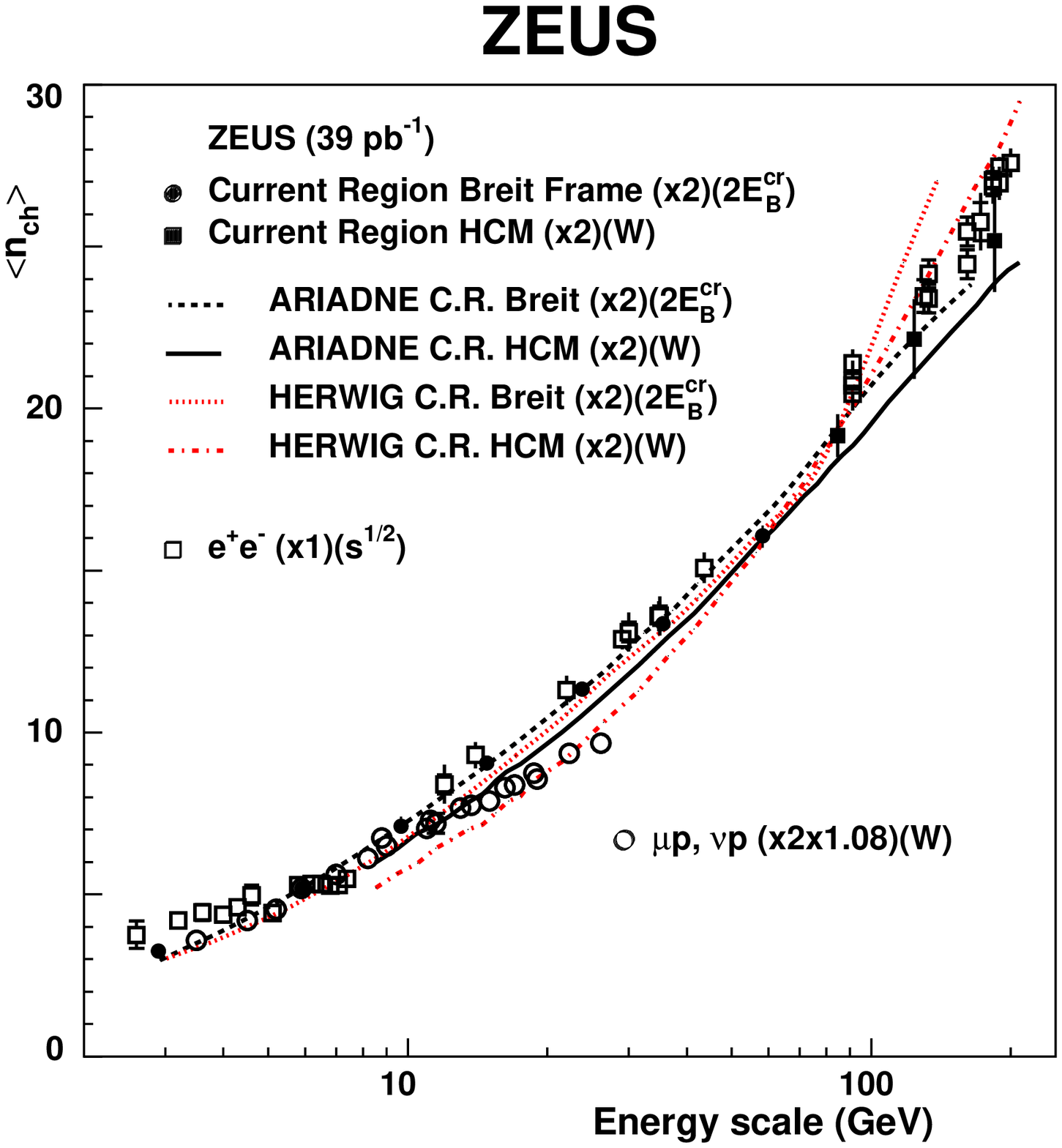}
\caption{The average charged multiplicity  ${\langle n_{\rm ch} \rangle} $
as a function of $2*E^{cr}_B$ and $W$ respectively for 
the Breit and  HCM frames.
Left: compared with the scale defined by  $Q$  and with MC predictions.
Right: multiplied by 2 compared with 
the measurements from $e^+e^-$~\cite{ee}  and the fixed target experiments
as well as with MC predictions.
} \label{obraz2}
\end{figure*}

\section{MULTIPLICITY OF  CHARGED HADRONS}

The data collected with the H1   detector
correspond to an integrated luminosity of
$44$~pb$^{-1}$~\cite{Aaron:2007ds} and with 
the ZEUS detector  to $39$~pb$^{-1}$~\cite{zeus,zeus99}.
The measurements of the average multiplicity, 
$\langle n_{\rm ch} \rangle$, 
presented in~Figs~\ref{obraz1}~and~~\ref{obraz2}
are performed in the current region
of the Breit frame  and of the HCM frame.
In the Breit frame the energy scale is 
related to the momentum of the scattered parton, $Q/2$.
Its hadronisation features are compared in Fig.~\ref{obraz1}
with  similar observables
measured in one hemisphere of  hadronic final states in $e^+e^-$
annihilation  described by  half of 
the centre-of-mass energy~$E^*$.
The large uncertainties for these ZEUS measurements
come from inclusion of the HERWIG hadronisation in systematic
uncertainties~\cite{zeus99}.
In Fig.~\ref{obraz2} an alternative energy scale 
to $Q$ is considered to be 
the available energy in the Breit current region $E^{cr}_B$ and
in the current region of the HCM frame $W$.
The $E^{cr}_B$ values are multiplied by 2 for consistency with Q.
%
In Figs~\ref{obraz1}~and~\ref{obraz2}
the MC models show good agreement with the data except the model  
with the soft colour interactions (SCI) (Fig.~\ref{obraz1}) 
and HERWIG at high energy scales~(Fig.~\ref{obraz2}).

Three energy scales $2*E^{cr}_B$, $W$ and $Q$ are considered.
As shown in Fig.~\ref{obraz2}~(left)~the average multiplicity
depends on the scale definition at scale below 10~GeV.  
This can be understood in terms
of higher order processes like BGF and migration of 
the emitted energy out of the Breit current region.
In Fig.~\ref{obraz1}  
the measurements are  below the $e^+e^-$ parameterisation
at the highest $Q^2$  in the vicinity of the $Z^0$ resonance.
The average charged multiplicities are also  compared 
in Fig.~\ref{obraz2}~(right)
with the $e^+e^-$ data and with the fixed target DIS 
measurements (see~\cite{zeus}).
The DIS data are scaled up by a factor 2 since they
are equivalent to one hemisphere of $e^+e^-$. 
All the  measurements exhibit approximately 
the same dependence of the average charged multiplicity
on the respective energy scale except the data from
the DIS fixed target experiments which deviate 
at energies above 15~GeV due to selection criteria.
More information on multiplicity characteristics 
can be found elsewhere~\cite{zeus}.

\section{SCALED MOMENTUM DISTRIBUTIONS}

\begin{figure*}[h]
\centering
\includegraphics[width=59mm]{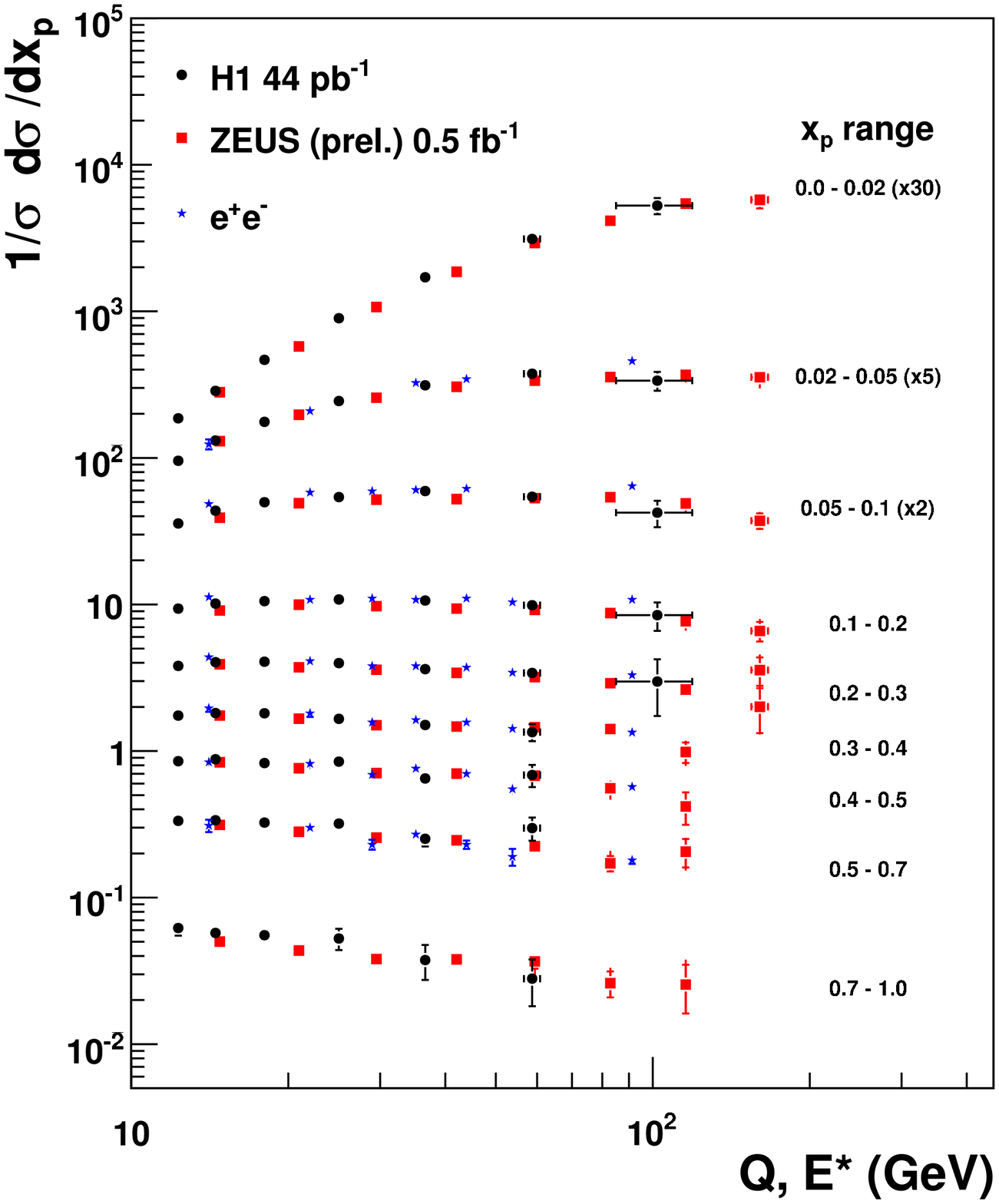}
\includegraphics[width=69mm]{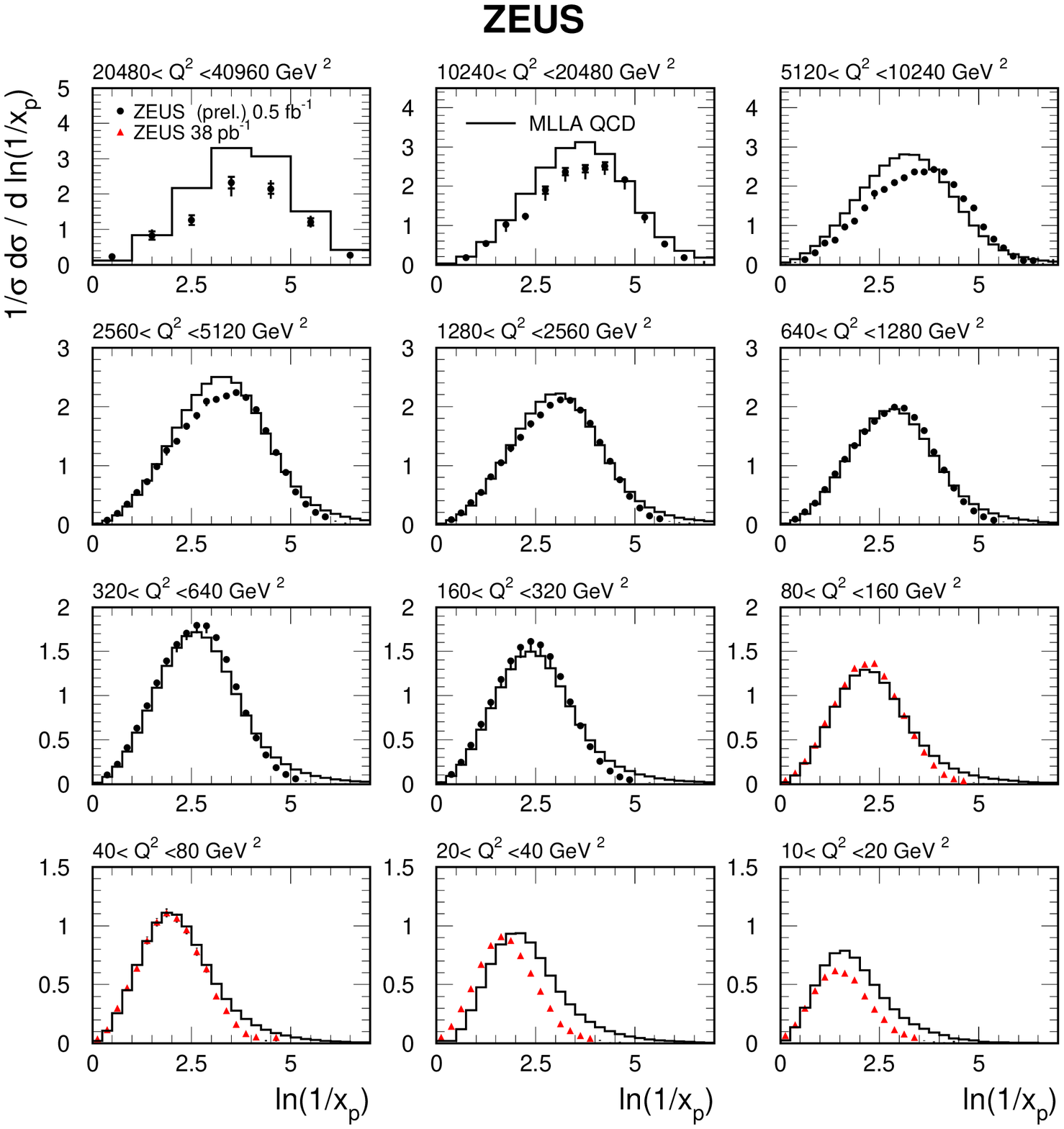}
\caption{Left: the charged particle distribution shown as 
 a function of $Q^2$ in $x_p$ bins; similar measurements 
from $e^+e^-$ are overlaid. Right: the normalized $\ln {(1/x_p)}$ 
distributions shown as the black dots for different $Q^2$ intervals;
the full line represents the MLLA+LPHD predictions.}  
\label{obraz3}
\end{figure*}

%
 The scaled momentum spectra have been studied 
 using an integrated luminosity of
 about 0.5~fb$^{-1}$  taken with the ZEUS detector 
 \cite{proc:zeusff:2007} and 44~pb$^{-1}$
 with the H1 detector~\cite{Aaron:2007ds}.  
The scaled particle momentum $x_p$ is defined as 
the momentum  of a particle measured in the Breit frame
scaled by the maximum momentum $Q/2$ of a quark in 
the current region.  Figure~\ref{obraz3}~(left) shows
the evolution of  scaled momentum distributions 
as a function of the available energy defined 
in $ep$ scattering by the exchanged boson virtuality 
$Q^2$.
There is  good agreement between different sets of $ep$ data. 
Scaling violations are observed.
The evolution is compared with the evolution of
the similar distributions  obtained from 
$e^+e^-$ experiments~\protect\cite{ee}.
%
The $ep$ data show similar behaviour to the $e^+e^-$  data
providing a rough demonstration of fragmentation 
universality.
However, a detailed comparison between $e^+e^-$  and $ep$ 
shows some discrepancies.
At high values of $Q^2$ approaching the $Z^0$ exchange region
and at  $x_p<0.1$ there are less charged particles observed
in the $ep$ data than in $e^+e^-$ annihilation.  
Similar observations  at $Q^2<100$~GeV$^2$
in previous measurements have been understood in terms
of higher order QCD processes  which deplete the current 
region in the Breit frame.

 The fragmentation function~(FF) represents the probability
 for a parton to fragment into a particular hadron carrying
 a fraction of the parton energy and incorporates the long distance,
 non-perturbative physics of the hadronisation process.
 Like structure functions,  the FF
 cannot be calculated in perturbative QCD, but can be evolved
 from a starting distribution at a defined energy scale
 using the DGLAP evolution~\cite{sovjnp:15:438} 
 equations.
 The fragmentation functions are universal and are the same
 in  $ep$ and $e^+e^-$.

The best way to exhibit the fragmentation features  are
hadron spectra as a function of $\ln {(1/x_p)}$ in  $Q^2$ 
intervals as shown in~Fig.~\ref{obraz3}~(right).
The spectra become softer with $Q^2$ increasing.
The main features of the data are 
reproduced by the MC predictions of ARIADNE and LEPTO (not show here)
except normalisation at the highest $Q^2$ values.
The modified leading log approximation (MLLA), presented 
also in Fig.~\ref{obraz3}~(right), describes
basic properties of particle production by multi-gluon
emission at leading order including colour coherence 
and gluon interference phenomena.
The  MLLA calculations include hypothesis of
local parton hadron duality (LPHD)~\cite{Azimov:1984np}
with parameters fitted to reproduce the $e^+e^-$ data.
The analytical MLLA+LPHD predictions describe the data at  
medium $Q^2$ but not at the lower and highest $Q^2$.
This can be partially explained by a significant migration
of particles from (or to) the current region of the Breit frame
due to the contribution of boson-gluon fusion processes.
As $Q^2$ increases the peaks are shifted more than expected towards
higher values of $\ln {(1/x_p)}$.  As the energy scale $Q$ increases
the coupling constant $\alpha_s$ decreases and amount of the soft
gluon radiation increases. This leads to a fast increase of the particle
density with small fractional momentum $x_p$.

\section{SUMMARY AND CONCLUSIONS}

The average charged multiplicity has
been  investigated in inclusive neutral current deep inelastic
$ep$ scattering in the kinematic range $Q^2 >25$~GeV$^2$ 
in terms of different energy scales. The virtuality of
the exchanged boson $Q$ and the scale $2*E^{cr}_B$ were used
in the current region of the Breit frame. 
In the current region of the HCM frame $W$ was used.
The  energy scales $2*E^{cr}_B$  and $W$ give better agreement with
measurements in $e^+e^-$ annihilation than $Q$.

The comparison of the scaled momentum distributions 
of  charged particles produced in $ep$ scattering with  
the same observables measured in $e^+e^-$ annihilation
support broadly the concept of quark fragmentation 
universality.  Some depletion of the scaled momentum distributions
at low $Q^2$ can be attributed 
to higher order QCD processes like BGF occurring in $ep$ but not in $e^+e^-$.
At high $Q^2$ approaching the $Z^0$ exchange region 
a significant depletion is also observed.
The $x_p$ distributions support 
the concept of quark fragmentation  universality in $ep$ scattering
and $e^+e^-$  annihilation.
The MLLA+LPHD calculations based on the $e^+e^-$ data do not  describe
the  distributions  in the entire range of the measured $Q^2$.


\newcommand{\etal}{et al.}
\newcommand{\coll}{Collaboration}

\end{document}